\documentclass{article}
\usepackage{spconf,amsmath,graphicx,booktabs,algorithm,algpseudocode}

\title{An End-to-end Approach for Lexical Stress Detection based on Transformer}
\name{Yong Ruan$^{1,2}$,Xiangdong Wang$^{1*}$, Hong Liu$^1$, Zhigang Ou$^3$,Yun Gao$^3$, Jianfeng Cheng$^3$ ,Yueliang Qian$^1$}
\address{
  $^1$Institute of Computing Technology, Chinese Academy of Sciences\\
  $^2$University of Chinese Academy of Sciences\\
  $^3$New Oriental Education and Technology Group \\
  \{ruanyong17g, xdwang, hliu\}@ict.ac.cn,\{ouzhigang,gaoyun10,chengjianfeng3\}@xdf.cn\},ylqian@ict.ac.cn}

\begin{document}
%
\maketitle
\begin{abstract}
The dominant automatic lexical stress detection method is to split the utterance into syllable segments using phoneme sequence and their time-aligned boundaries. Then we extract features from syllable to use classification method to classify the lexical stress. However, we can't get very accurate time boundaries of each phoneme and we have to design some features in the syllable segments to classify the lexical stress. Therefore, we propose a end-to-end approach using sequence to sequence model of transformer to estimate lexical stress. For this, we train transformer model using feature sequence of audio and their phoneme sequence with lexical stress marks. During the recognition process, the recognized phoneme sequence is restricted according to the original standard phoneme sequence without lexical stress marks, but the lexical stress mark of each phoneme is not limited. We train the model in different subset of Librispeech and do lexical stress recognition in TIMIT and L2-ARCTIC dataset. For all subsets, the end-to-end model will perform better than the syllable segments classification method. Our method can achieve a 6.36\% phoneme error rate on the TIMIT dataset, which exceeds the 7.2\% error rate in other studies.
\end{abstract} 
\begin{keywords}
Lexical stress detection, computer assisted language learning, end-to-end, transformer
\end{keywords}

\section{Introduction}
\label{sec:intro}
Lexical stress is associated with the prominent syllable of a word. In English, at least one syllable in a polysyllabic word should be stressed relative to other syllables. Lexical stress plays a very important role in the understanding and perception of speech. Modifying the lexical stress of a word may change the meaning of the word. For example, the word "record" can be a noun or a verb when the stress is located differently. The position of the lexical stress of a word is unpredictable, but it is customary, and each meaning of a word has a unique corresponding lexical stress. Therefore, the English pronunciation dictionary contains the position of the lexical stress in addition to the pronunciation of the word \cite{shahin2016automatic}.

In the process of learning a second language such as English, the mastery of lexical stress is very important. Computer-aided pronunciation assessment systems need to identify where the words are stressed to help non-native language learners learn the language. The systems need to evaluate the learner's pronunciation quality and provide the learner with feedback on the lexical stress pronunciation, so as to help the speaker to better grasp the lexical pronunciation. Therefore, the automatic evaluation technology of lexical stress is an important part of evaluating the pronunciation quality of the speaker, and plays an important role in applications such as computer-aided language learning systems \cite{chandel2007sensei}.

There has been much research work on the assessment of lexical stress, especially for native and non-native English speakers. Most of the current methods extract features for a syllables, and then the lexical stress detection task is considered as a classification problem to determine whether a syllable should be stressed or not. Several temporal features such as loudness, duration, pitch \cite{chen2010automatic,li2013lexical} are commonly used currently. In \cite{chen2007using}, the authors used a combination of temporal features and spectral features to obtain better results. In \cite{yarra2017automatic}, the authors propose a new feature contour based on sonority and find an optimal set of sub-bands to improve the accuracy of lexical stress detection.

Based on the features extracted, many classification methods are used to detect lexical stress, such as support Vector Machines(SVM) \cite{chen2010automatic,zhao2011automatic}, Hidden Markov Models(HMM) \cite{stehwien2019convolutional} and Maximum Entropy\cite{sridhar2007exploiting}. Recently, the classification method of deep learning is also applied to the lexical stress classification. In \cite{li2013lexical}, the authors use the deep belief network for lexical stress classification, which has an accuracy improvement of 8$\%$ compared to the traditional Gaussian Mixture Model. In \cite{shahin2016automatic}, the authors compared the error rates of deep-connected neural networks and deep convolutional Neural Networks(CNNs) for lexical stress classification, and it is reported that the CNN model achieves better results. In \cite{sridhar2007exploiting}, the authors found that CNN can yield good performance in lexical stress detection using low-level acoustic features alone, without any high level features such as duration of phonemes and syllables. 

The above method mainly uses the features in the syllables to perform lexical stress classification. In \cite{yarra2019comparison}, the author combines the acoustic based features with the text content based features, and obtains better results than only using the acoustic based features.

For the methods based on syllable classification, temporal boundaries need to be determined. Current approaches commonly adopts a two-stage process, where the boundaries of syllables are first determined by a forced-alignment between the speech and the text (which can be transformed into a sequence of syllables using a pronunciation dictionary) using a trained acoustic model and then features are extracted from the syllables and classifiers are applied. Obviously, the time boundaries obtained by method may be with errors and the errors will further affect the performance of classification. More importantly, stress in speech is actually a global characteristic of the whole utterance and the stress of a certain syllable depends not only on the acoustic feature of itself, but also on the contextual information (e. g., energy, frequency) of other syllables or even the entire utterance. Although some methods extract features from adjacent syllables to incorporate more context, the contextual information they use is still very limited, with a serious lack of long-distance and global information within the utterance. 

In \cite{ramanathi2019asr}, the authors propose to estimate stress markings in automatic speech recognition (ASR) framework involving finite-state-transducer (FST) without using annotated stress markings and segmental information. First, they train the ASR system with native English data along with pronunciation lexicon containing canonical stress markings. In the decoding phase, the phoneme sequence with different type of lexical stress positions derived from the phoneme sequence of the speech is used to establish the decoding path. The corresponding lexical stress classification result can be obtained by selecting the lexical phoneme path with the highest probability. Although this method avoids error cascade by performing a one-pass decoding, the contextual information it used is still limited. The process of decoding only uses acoustic features of current phonemes for classifying and the adjacent phonemes considered are also limited.

In this paper, with insight of long-distance or global contextual information needed in lexical stress detection, we propose a sequence-to-sequence approach for lexical stress detection. The encoder-decoder structure of Transformer \cite{vaswani2017attention} is used to translate the feature sequence of the audio to a phoneme sequence with lexical stress marks directly. During the training process, the model is trained according to the feature sequence of the audio and the corresponding phoneme sequence with lexical stress marks. During the detection process, the recognized phoneme sequence is restricted by the  phoneme sequence of uttereance, but the lexical stress mark of each phoneme is not limited. For example, for the word "predict", its phoneme sequence is P R IH0 D IH1 K T, where IH0 and IH1 indicate that the corresponding phonemes (IH) are not stressed (0) and stressed (1), respectively. In the end-to-end decoding process, the restriction if that the utterance must be decoded in the phoneme order of P R IH D IH K T , but for each IH, it can be either IH0 or IH1. Finally, the lexical stress of each phoneme is determined based on the most probable phoneme sequence (with the greatest probability) with lexical stress marks. This method does not require the boundaries of syllables and does not require the use of manually designed features, nor does it require training of acoustic models. Most importantly, the sequence-to-sequence scheme it used and the self-attention modules in Transformer make full use of the global contextual information of the utterance and can bring better results compared to traditional methods, which is proved by our experiments. 

\section{Dataset}
\label{sec:format}
We mainly use three Datasets, Librispeech\cite{panayotov2015librispeech} is used for training and TIMIT\cite{garofolo1993darpa} and L2-ARCTIC\cite{zhao2018l2} are used for testing.

LibriSpeech ASR corpus is a large corpus containing about 1000 hours of English speech. These data are from audio books from the LibriVox project. We divided the Librispeech data set into three subsets of 100 hours, 460 hours, and 960 hours. Training experiments were conducted on these three subsets.In the test phase, the TIMIT dataset and L2-ARCTIC dataset are used as test sets.

The TIMIT data set contains a total of 6300 utterances of sentences. Each of the 630 speakers from the eight major dialect regions of the United States read 10 given sentences, all of which are manually segmented and labeled at the phoneme level.
 
The L2-ARCTIC corpus is a corpus of non-native English speech and is intended for research in voice conversion, accent conversion, and mispronunciation detection. In total, the corpus contains 22,339 utterances from twenty (20) non-native speakers with a balanced gender and L1 distribution. Most speakers recorded the full CMU ARCTIC set. Human annotators manually examined 2,999 utterances, so we use this part as our test dataset.
 
We use the Carnegie Mellon University(CMU) pronunciation dictionary as a unified pronunciation dictionary with a total of 39 phonemes. The unified pronunciation dictionary is used to convert the sentences corresponding to the audio into phoneme sequences.

\section{Method}
\label{sec:format}

\begin{figure}[htb]

\begin{minipage}[b]{1.0\linewidth}
  \centering
  \centerline{\includegraphics[width=8.5cm]{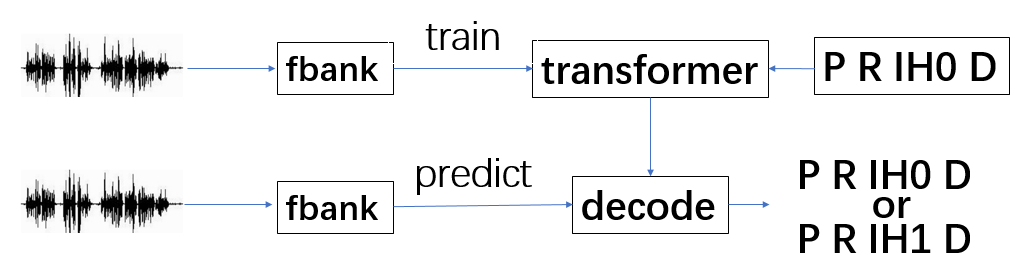}}
  \centerline{\textbf{Fig. 1}.Procedure of the proposed method for lexical stress detection}\medskip
\end{minipage}

\end{figure}

Fig. 1 shows the procedure of the proposed method for lexical stress recognition with an example of P R IH0 D. We use frame length of 25ms and frame shifting of 10ms to extract the 80-dimensional mel-bank magnitudes. During training, using the feature sequence of the  audio and the corresponding sequences with stress marks (P R IH0 D), we train a sequence-to-sequence model based on the Transformer including the encoder and the decoder for recognition of phonemes with stress marks. In the recognition phase, the original recognition method is modified With the restriction of the given sequence of phonemes, the most probable sequence of phonemes with stress marks is searched among all possible paths. And the stress of each syllable can be determined according to the sequence. With the restriction of the given sequence of phonemes, the most probable sequence of phonemes with stress marks is searched among all possible paths. And the stress of each syllable can be determined according to the sequence. For example, according to the phoneme sequence of P R IH D, since the IH phoneme may have lexical stress or may not, there may be two possible  of P R IH0 D and P R IH1 D. After the decoding, IH can be determined to be stressed or not stressed.

\subsection{Model}
\label{ssec:subhead}

\begin{figure}[htb]

\begin{minipage}[b]{1.0\linewidth}
  \centering
  \centerline{\includegraphics[width=8.5cm]{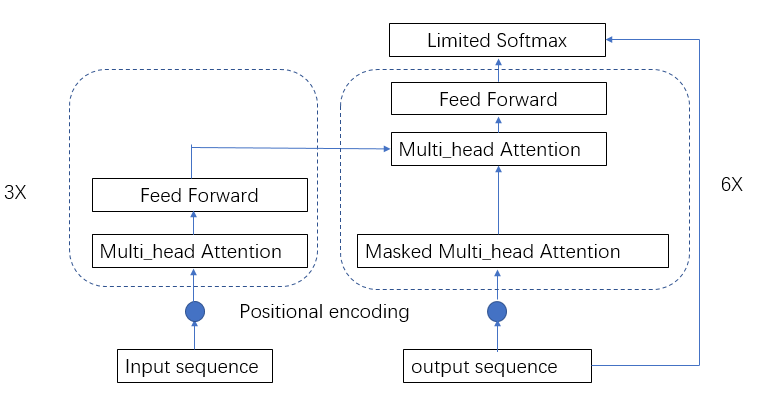}}
  \centerline{\textbf{Fig. 2}.The Transformer architecture}\medskip
\end{minipage}

\end{figure}

As Fig. 2, we use the Transfomer [12] model that relies on the self-attention mechanism. For an input sequence ($x_{1}$,...,$x_{n}$), the encoder converts it into a continuous expression sequence z=($z_{1}$,...,$z_{n}$). According to this expression sequence, the decoder generates an output symbol on a time-by-time basis. When an output symbol of a certain period of time is generated, the output symbol before the time period is taken as an input, and finally an output symbol sequence ($y_{1}$,...,$y_{n}$) is generated. Both the encoder and the decoder use some identical layers composed of self-attention and fully connected layers. Both the encoder and the decoder are stacked from several identical layers, each layer including a multi-head and point-by-point fully connected network. Each layer in the decoder also includes a multi-head attention according to the output of the decoder.

\subsection{Train}
\label{ssec:subhead}

We train this model on the Librispeech dataset. We use the Adam optimizer and the relevant parameters are chosen to be $\beta$ 1=0.9, $\beta$ 2=0.997, $\epsilon$ =$10^{-9}$. The learning rate is set to 0.15 and uses 0.1 dropout and 0.1 label smoothing. We train up to 500k steps, and use early-stopping based on the performance on the verification set.

\subsection{Recognition}
\label{ssec:subhead}

The model trained in the training stage is used in the recognition process, but unlike the original recognition method, the recognition paths are restricted in the recognition process. In original recognition, greedy search is used, and at each time, the most probable output symbol is chosen. But in lexical stress recognition, the decoding algorithm makes limited choices based on the phoneme sequence of the utterance at each moment. For example, for PR IH0 D, its phoneme sequence without lexical stress mark is PR IH D, which is limited to P and R at the first and second period, and for the third period, because IH has two choices of IH0 and IH1, Therefore, the probability of selection is between IH0 and IH1. The output of the fourth period will be limited to D. At this period, the whole phoneme sequence is output and the recognition ends. For IH phonemes, the lexical stress selection of the IH phoneme is determined based on the probability of selection during the recognition process.

\begin{algorithm}[htb]
\caption{:Decoding}
\label{alg:Framwork}

\textbf{Inputs:} x,y

\textbf{Outputs:} p

predict=[]

encode-input = encode(x)

for i in length(y):

\quad	p = y[i]

\quad	logits = decode(encode-input,y,predict)

\quad	y-predict = argmax(logits in p0 ,p1 and p2)

\quad	predict.append(y-predict)

return predict
\end{algorithm}

As shown by the pseudo code, in the decoding phase, the feature sequence x of an utterance and the phoneme sequence y of the corresponding text are input. The feature sequence x is first encoded using an encoder. Then, the output of the encoder is decoded according to the length of the phoneme sequence of the corresponding text. Each time, the decoder uses the output of the encoder, phoneme p at the current position of the phoneme sequence, the prediction result of the encoder before this time to obtain the probability of each phoneme with the lexical stress mark. According to the phoneme p of the current position, the predicted phoneme with the highest probability is selected in the set of lexical stress marks of the phoneme p, and the phoneme is added to the prediction result and then predicted at the next period. Finally, the predicted result of the specified length is obtained.

\section{Experiment}
\label{sec:pagestyle}

In the experiment, the Librispeech data set was used as the training set of the model. We use the method of classifying syllable features using a classifier as the baseline. We use the method in \cite{shahin2016automatic} to first train an alignment model using the Librispeech dataset, and then use the alignment model to align the audio features with the audio phoneme sequence to get the phoneme boundaries, and then get the syllable boundaries. 27-dimensional mel-bank features are extracted within the syllable range and  
we use a window of 3 consecutive syllables to classify. We use two layers of CNN and one layer of DNN to obtain the lexical stress classification result.

We also use the end-to-end model of the transformer for lexical stress detection. The end-to-end model is first trained on the Librispeech dataset based on the extracted mel-bank feature sequences and the phoneme sequences with lexical stress marks. The recognition phase performs decoding of the restricted phoneme sequence. The decoding at each moment limits the category of phoneme without limiting the type of lexical stress marks, and finally the lexical stress pronunciation of each phoneme is obtained from the phoneme sequence with lexical stress marks of the maximum probability.

 The phoneme error rate of the lexical stress detection is used as the evaluation metric, and only the lexical stress phonemes in multi-syllable words are counted in the statistical process. The experimental results are shown in Table 1 and Table 2.

\begin{table}[h!]
  \centering
  \caption{Results in TIMIT}
  \label{tab:table1}
  \begin{tabular}{ccc}
    \toprule
    dataset & Segmented syllable classification & transformer\\
    \midrule
    100 hours & 13.92\% & 10.62\%\\
    460 hours & 10.12\% & 7.92\%\\
    960 hours & 10.31\% & 6.36\%\\
    \bottomrule
  \end{tabular}
\end{table}

As shown in Table 1, for the end-to-end model, as amount of data increases, the error rate decreases gradually. Because the end-to-end model needs to learn related phoneme sequence information and phoneme alignment information in one model at the same time, the amount of data required is relatively large. Therefore, as the data size increases, the phoneme error rate gradually decreases. For the segmented syllable classification experiment, the 460-hour data error rate is lower than the 100-hour data, but when increasing the data size to 960 hours, the error rate does not decrease. Because the 100-hour data is relatively pure data, the distribution of TIMIT data sets is quite different. For 460 hours of data, some data with a slightly lower quality is added, and the amount of data is increased, which will get better results. With 460 hours of data sufficient, there will be no significant increase in the amount of data that continues to increase. For all subsets of 100 hours, 460 hours, and 960 hours, the end-to-end model performs better than the segmented syllable classification system. Our method can achieve a 6.36\% phoneme error rate on the TIMIT dataset, which exceeds the 7.2\% error rate of the method in the TIMIT dataset in \cite{shahin2016automatic}.

\begin{table}[h!]
  \centering
  \caption{Results in L2-ARCTIC}
  \label{tab:table1}
  \begin{tabular}{ccc}
    \toprule
    dataset & Segmented syllable classification & Transformer\\
    \midrule
    100 hours & 15.3\% & 14.2\%\\
    460 hours & 15.7\% & 8.7\%\\
    960 hours & 15.8\% & 8.8\%\\
    \bottomrule
  \end{tabular}
\end{table}

For the L2-ARCTIC dataset, the transformer model also performs well. For the segmented syllable classification experiment, there is no particularly significant decrease in phoneme error rate as the train data size increases.But for the transformer model, the phoneme error rate decreases significantly when the dataset size increases to 460 hours. And it has no improvement when its increases continuely. This may be because that speech in the L2-ARCTIC are non-native, but it need to be further confirmed in future work. For all subsets, the transformer model performs better than syllable based features method, which proves the superiority of the transformer method.

\section{Conclusions}
\label{sec:typestyle}

In lexical stress detection, a method of classifying using a classifier based on a syllable extraction feature or a method of constructing and decoding an FST based on a frame of speech recognition is generally employed. We propose to use the end-to-end Transformer model to train audio features directly into the phoneme sequence recognition of lexical stress marks. By limiting the phoneme sequence in the phoneme recognition process, we find the most likely lexical stress sequences to obtain the lexical stress recognition results. The method is better than the method based on syllable feature extraction using classifiers, and the result of the phoneme error rate of 6.36\% on the TIMIT data set is better than the best result on the previous TIMIT data set. The end-to-end method is performing well in TIMIT dataset and L2-ARCTIC dataset and this method is meaningful to continue to explore.

\clearpage

\bibliographystyle{IEEEbib}
\bibliography{strings,refs}

\end{document}